\documentclass[preprint,aps,showpacs,preprintnumbers,amsmath,amssymb,superscriptaddress,showkeys]{revtex4}

\usepackage{epsfig}

\usepackage{graphics}
\usepackage{dcolumn}
\usepackage{bm}

\begin{document}


\title{Unzipping of DNA under the influence of external fields}

\author{A. E. Bergues-Pupo}

\affiliation{Departamento de F\'isica, Universidad de Oriente, 90500 Santiago de Cuba, Cuba}

\affiliation{Dpto. de F\'isica de la Materia Condensada,
Universidad de Zaragoza. 50009 Zaragoza, Spain}

\author{J. M. Bergues}

\affiliation{Escuela Polit\'{e}cnica Superior y Facultad de Ciencias de la Salud, Universidad San Jorge, 50830 Villanueva de G\'{a}llego, Zaragoza, Spain}

\author{F. Falo}

\email{fff@unizar.es}

\affiliation{Dpto. de F\'isica de la Materia Condensada,
Universidad de Zaragoza. 50009 Zaragoza, Spain}

\affiliation{Instituto de Biocomputaci\'on y F\'isica de Sistemas Complejos (BIFI), Universidad de Zaragoza, 50009 Zaragoza, Spain}

\date{\today}

\begin{abstract}
We study the features on the unzipping process of a modified version of the Peyrard-Bishop-Dauxois model. We show that the inclusion of a barrier in the on-site potential allows for the existence of stable domain wall solutions between open and closed regions of the DNA chain. We analyze the linear stability of such solutions and study their relevance in the dynamical behavior of DNA under ac forces. 

\end{abstract}


\maketitle

\section{Introduction}

The study of the energy landscape of biomolecules is possible, among other methods, through their mechanical response under the action of an external force. These studies are important because, in many cellular process like transcription or replication, DNA is under the action of different mechanical stresses. One of the experiments that shows these potentialities is the mechanical unzipping of DNA \cite{Bockelmann1997,Bockelmann2002,Danilowicz2003}. It consists of separating the DNA molecule double helix by pulling apart one strand from the other one by using an atomic force microscope, optical or magnetic tweezers.  The obtained signal can give information of base pair (bp) content of a sequence \cite{Bockelmann1997,Bockelmann2002,Danilowicz2003} or may be used to estimate interaction energies of bps through the chain \cite{Ritort2010}.

Special attention has been given in the last years to the influence of external ac fields on the dynamic response of DNA  \cite{Alex_PRA,Alex_PRE,Ana,Alex_PLoSCB,Bruan_2004,Ritort2012,Kumar2013}. From one side, the possible effects of THz fields have been discussed \cite{Alex_PRA,Alex_PRE,Alex_PLoSCB,Ana}. The argument for such an influence resides on the fact that weak bonding energies of the molecule, e.g. hydrogen bonds between nucleotides of a bp, are in the THz frequency range and thus resonant effects could be expected. On the other side, ac mechanical forces acting during single molecule experiments have been addressed \cite{Bruan_2004,Ritort2012}. In \cite{Bruan_2004} the use of a periodic driving protocol is proposed for the folding and unfolding of a protein and it is obtained a better resolution in the reconstruction of the free energy landscape of the molecule. On the other hand, stochastic resonance and resonant activation were observed during the folding unfolding of short DNA hairpins \cite{Ritort2012}. The resonant frequencies obtained in this case match the hopping dynamics of the system rather than the natural frequencies and may be used to estimate kinetics rates. The idea of the use of mechanical alternate forces is also supported by the fact that many process occur in a periodic way due to the periodic energy consumption inside the cell \cite{Kumar2013}.


The use of simple physical models may help in the understanding of such kind of processes. In fact, some of the mechanical and thermal properties of DNA, and other biomolecules, can be modeled at a mesoscopic level. One of the most studied models at this scale is the Peyrard-Bishop-Dauxois (PBD) model \citep{Peyrad1993}. The only degree of freedom relevant for this model is the inter-strand separation. Despite its simplicity, it contains the main ingredients to describe the phenomena mentioned above. It was initially proposed for explaining the melting transition of DNA, i.e., the separation of the double strand upon heating, but it has been also used to explain other phenomena as DNA mechanical unzipping \cite{Santiago2005,Voulgarakis2005,Voulgarakis2006,Singh2013}, microscopic mechanical flexibility of DNA \cite{Webber2009} and the influence of external ac fields on the DNA dynamics \cite{Alex_PRA,Alex_PRE,Ana}.

In a recent work \cite{Ana}, we studied the influence of an ac external field on the melting transition and bubble formation of homogeneous and heterogeneous DNA sequences. We used a modified PBD model that includes a solvation barrier that takes into account the interaction of the molecule, once opened, with the solvent. This term has been used in previous works in order to provide results like bubble lifetime and melting width closer to those of experimental data \cite{Weber2006,Peyrard2009,Tapia} and to model the unzipping process under different salt concentrations of the solution \cite{Singh2013}. We found that the external field influences resonantly the DNA dynamics at certain frequency values, corresponding approximately to the natural oscillation frequencies of AT and CG bps inside the Morse potential. Consequently the response of the system distinguishes the AT-rich regions from CG-rich regions. Interestingly, if the barrier is not included, this influence is almost independent of the frequency.

The introduction of the barrier term leads also to new features at the basic level. For instance, with the standard Morse potential (without including the barrier) a domain wall solution of the form $Y(x)=\ln  [1+e^{\sqrt{2/S}(x-x_0)}]$ can be obtained \cite{Peyrad_2004},  where $Y$ represents the separation of bases at position $x$ and $S$ set the wall width (and it is dependent on model parameters).This solution describes a configuration where one part of the molecule ($x<x_0$) is closed and the other one ($x\gg x_0$) is opened and the bp separation grows linearly with space. However, this solution, for the standard PBD model, is unstable \cite{Peyrad_2004}. We show here that with the barrier term a stable domain wall solution can be found and thus, different oscillation modes corresponding to the different opening states of the chain can be obtained.

In this paper, we will focus on the analysis of the unzipping process. We show that the inclusion of the barrier on the Morse potential allows us to obtain a stable domain wall solution for different values of the unzipping force. The stability analysis of this solution allows us to establish the frequency interval where resonant mechanisms can be found and also to estimate the escape rates for the unzipping. With these elements we finally study the combination of thermal noise and an external ac field in the unzipping process. The paper is developed as follows: model and methods are shown in section II. The linear stability analysis of the domain wall solution is presented in section III. The estimation of the escape rates is made in section IV and the influence of the ac field in the unzipping process is presented in section V.

\section{Model and Methods}

We use the PBD model with the inclusion of a Gaussian barrier as in \cite{Ana,Tapia}.  Similar models with other expressions for the barrier were proposed in \cite{Weber2006,Peyrard2009}. The 1d Hamiltonian, in terms of the separation of bp $y_n$, is given by the following expression:

\begin{equation}
H=\sum_n \left[ \frac{p_n^2}{2m}+V(y_n)+W(y_n,y_{n-1}) \right],
\label{eq:ham}
\end{equation}

where the first term corresponds to the kinetic energy ($p_n = m dy_n/dt$ is the momentum of the $nth$ bp and $m$ its reduced mass) and the potential energy is given by the two other contributions: the on-site potential $V(y_n)$, that describes the interaction of bases belonging to the same pair; and the stacking interaction $W(y_n,y_{n-1})$, that gives the interaction of consecutive bps. $V(y_n)$ is defined as the sum of the Morse potential and a Gaussian barrier:

\begin{equation}
V(y_n)=D ({\rm e}^{-\alpha y_n}-1)^2 + G {\rm e}^{-(y_n-y_0)^2/b}.
\label{eq:Vmod}
\end{equation}

\begin{figure}
\includegraphics[width=10truecm]{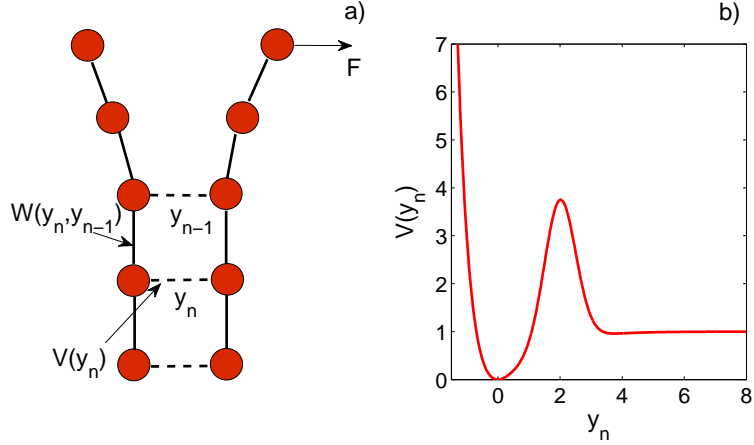}
\caption{(Color online) a) Scheme of the unzipping process. b) Morse potential with the Gaussian barrier.}
\label{fig:modelo}
\end{figure}

$D$ is the bp dissociation energy and $\alpha$ sets the width of the potential well. The Gaussian barrier models the entropic barrier that bps have to overcome to open and close again. The origin of this term comes from the fact that when the hydrogen bonds are broken, the bases can form new bonds with the solvent molecules and thus there must be an energetic cost to close them \cite{Weber2006,Tapia}. The parameters $G$, $y_0$ and $b$ are the barrier height, position and width respectively.
The term $W(y_n,y_{n-1})$ accounts for the stacking interactions:

\begin{equation}
W(y_n,y_{n-1})=\frac{1}{2}K(1+\rho{\rm e}^{-\delta(y_n+y_{n-1})})(y_n-y_{n-1})^2,
\end{equation}
$K$ is the coupling constant, $\rho$ sets the anharmonic character  of the stacking interaction and $\delta$ sets the scale length for this behavior.
To simulate the unzipping under constant force, we added a term $V_f=-Fy_1$ to the first bp of the chain. A schematic view of the model and the potential $V(y_n)$ is depicted in figure \ref{fig:modelo}.

The values of the potential parameters are the same of those used in \cite{Ana,Tapia} for a homogeneous AT chain: $D=0.05185$ eV and $\alpha=4$ $\AA^{-1}$, for the Morse potential; $G=3D$, $y_0=2/\alpha$ and $b=1/2\alpha^2$ for the Gaussian barrier; and $K=0.03$ eV$\AA^{-2}$, $\delta=0.8$  $\AA^{-1}$ and $\rho=3$, for the stacking interaction. For simplicity, we use dimensionless units: length is given in units of $\alpha^{-1}$; mass in units of the nucleotide mass (300 uma) and energy in units of $D$. With these transformations we get $F_u = 332$  pN, $\omega_u = 5.15$  rad/ps (0.82 THz) (and $T_u=602$  K, for the force, frequency and temperature units, respectively.

Given the shape of the potential $V(y_n)$, a stable domain wall solution can be obtained for different values of $F$. Beyond a critical value $F_d$, called the depinning force, the domain wall displaces from its equilibrium position and no static solution can be found. We carry out the linear stability analysis of this solution. For this, the configuration $\{y_n^{eq}\}$ satisfying the equilibrium conditions has to fulfill:

\begin{equation}
\frac{\partial (V(y_n,F)+W(y_n,y_{n-1}) +W(y_{n+1},y_n))}{\partial y_n}=0.
\label{eq:eq_cond}
\end{equation}
where $V(y_n,F)=V(y_n)-Fy_1\delta_{1n}$. To solve the nonlinear equation system \ref{eq:eq_cond}, it is necessary to propose an initial guess for the solution. As we are interested in a domain wall solution, we start from a configuration with half of the chain at $y_n<y_0$ (closed state) and the other one at $y_n>y_0$ (open state). Using this initial guess, the nonlinear equation system \ref{eq:eq_cond} is solved. Once a solution is reached, we proceed to analyze its linear stability i.e. its behavior under small perturbation. To do that we build the Jacobian matrix of forces on each particle or, equivalently, the Hessian matrix of the potential energy \cite{Strogatz}. The eigenvalues $\omega^2_i$  obtained by diagonalizing the Hessian matrix of the system provide the signatures of the equilibrium solution. If the solution is stable, all eigenvalues have to be positive.

An unstable equilibrium solution can also be obtained for equation system \ref{eq:eq_cond} by using the following procedure: first, the stable equilibrium configuration $y_n^{eq}$ is determined. Then, we add a small displacement to the bp on the close state that is closer to the open state ($y_i^{eq}\rightarrow y_i^{eq}+dy$). The system of equations \ref{eq:eq_cond} is solved again with the initial ansatz given by $y_n=y_n^{eq}$, $n\neq i$ and $y_i^{eq}$ as a parameter, i.e., the system has $N-1$ equations. The energy of the obtained configuration $E_1$ is computed. The same procedure is repeated by displacing $y_i^{eq}$ further, i.e., $y_i^{eq}\rightarrow y_i^{eq}+sdy$ ($s=1,2,...$), solving the system equations for the remaining $y$ and calculating the energy $E_s$ for the obtained configuration. The procedure stops when a maximum value of the energy is reached. The configuration at this energy value corresponds to the unstable solution of the system. The eigenvalues of the stability matrix are all positive except one which corresponds to the unstable direction i.e. a saddle point in the energy landscape. 

If thermal noise or an external ac field are included, the Langevin equations of the motion are integrated numerically:

\begin{eqnarray}
m\frac{\partial^2 y_n}{\partial t^2}+m\gamma\frac{\partial y_n}{\partial t} &=&-\frac{\partial\left [W(y_{n},y_{n-1})+W(y_{n+1},y_n)\right]}{\partial y_n}  \nonumber \\ & & -\frac{\partial V}{\partial y_n} +\xi_n(t) + A cos(\omega t),
\label{eq:lang_dyn_cont}
\end{eqnarray}
where $m$ is the reduced mass of the bp, $\gamma$ is the effective damping of the system, $T$ is the temperature, $A$ and $\omega$ are the amplitude and frequency field respectively. $\xi(t)$ is a white Gaussian noise with $\langle\xi_n(t)\rangle=0$ and $\langle\xi_n(t)\xi_k(t')\rangle=2m\gamma k_BT\delta_{nk}\delta(t-t')$. To integrate equation system \ref{eq:lang_dyn_cont} we use a stochastic Runge-Kutta algorithm \cite{sde1,sde2}.

\section{Results}

\subsection{Linear Stability Analysis of the domain wall solution}

We use a homogeneous AT chain of $N=10$ bps. One end has fixed boundary conditions, i.e., $y_{N+1}=0$ and the other one, where the force is applied, is free. We obtain the stable domain wall solution by solving \ref{eq:eq_cond} for different values of the constant force. The set of nonlinear equations  is integrated numerically within Matlab with the trust-region-dogleg method. The initial guess for the solution is set with half of the chain at $y_i = 5, i=1,..,5$ (open state) and the other one at $y_i = 0, i=6,..,10$ (closed state). We start from $F=0$ and then increase the force until no equilibrium configuration can be found. The force value at this point is defined as $F_d$. The unstable solution is also obtained by following the method depicted in the previous section.

\begin{figure}
\includegraphics[width=10truecm]{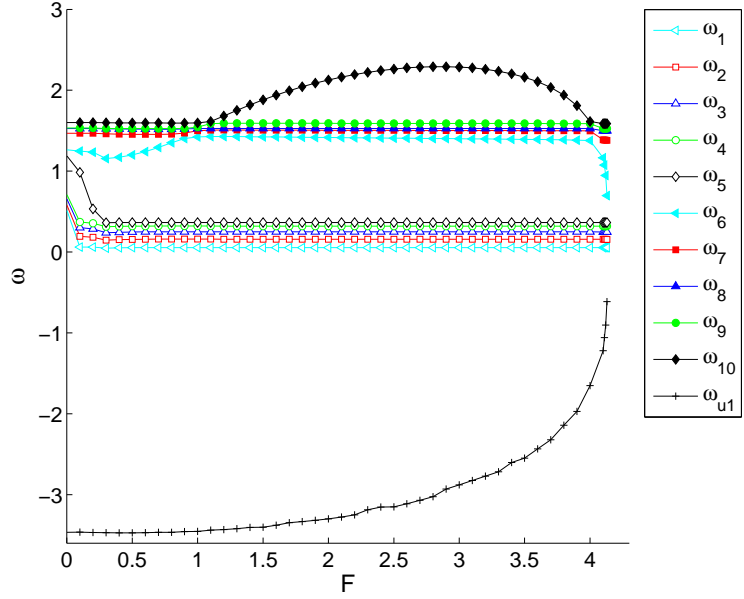}
\caption{(Color online) Frequency spectrum at different forces. $\omega_{u1}$ is the negative eigenvalue of the unstable solution.}
\label{fig:eigenvalues}
\end{figure}

The eigenvalues $\omega_i$ for the stable solution are shown in figure \ref{fig:eigenvalues}. The negative eigenvalue $\omega_{u1}$ for the unstable solution is also displayed.

\begin{figure} [bottom right]
\includegraphics[width=10truecm]{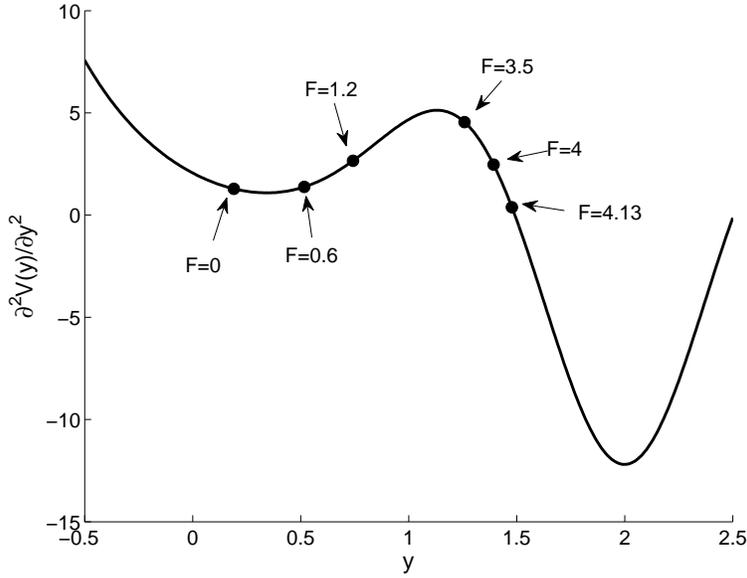}
\caption{Second derivative of the Morse potential and $6^ {th} bp$ positions at different forces.}
\label{fig:potential}
\end{figure}

Two frequency bands are identified: one at $0<\omega<0.5$ and the other one at $1.5<\omega<1.7$, corresponding to the open and closed part of the chain, respectively. The position of these bands may be estimated analytically using the dispersion relations for the linear excitations inside the Morse potential (closed chain), $m\omega^2\approx2D\alpha^2+2K(1+\rho)[1-\cos(\pi n/N)]$; and for a free Gaussian chain (open chain), $m\omega^2=2K[1-\cos(\pi n/N)]$ \cite{Tapia}. Thus, the lower frequency band corresponds to the open chain while the upper one to the closed chain.

One additional mode, corresponding to the domain wall between the open and the closed sides, is also observed. With our model parameters, the domain wall contains only one bp (the $6^{th}$ bp). The location of this mode depends on the value of the applied force. At $F<1.1$ and $F>4.1$ the mode is located between the lower and the upper frequency band while at $1.1<F<4.1$ is over the upper band. This may be explained because the second derivative of the Morse potential changes with position and $\omega ^2$ is directly proportional to it (see figure \ref{fig:potential}). When $F\rightarrow F_d$, the intermediate mode goes to the upper limit of the lower band. At $F=0$, frequency values corresponding to the open part of the chain are larger than those obtained for $F>0$. This is due to the presence of a local minimum beyond the barrier at $y\approx 3.75$. When the force is applied, these modes approach zero indicating that bps are displaced away from that minimum. We have obtained $F_d\approx 4.13$, which corresponds to a force of $1371$  pN. This force value is much larger than the mean force reported for DNA unzipping ($\approx 15$  pN) \cite{Bockelmann1997}. However, we cannot compare these numbers since $F_d$ is defined at zero temperature and no thermal effects are included. The real experiments are carried out at temperatures where thermal fluctuations help the opening of a bp. Thus, thermal fluctuations can reduce the unzipping forces as we discuss below.

\begin{figure} [tr]
\includegraphics[width=10truecm]{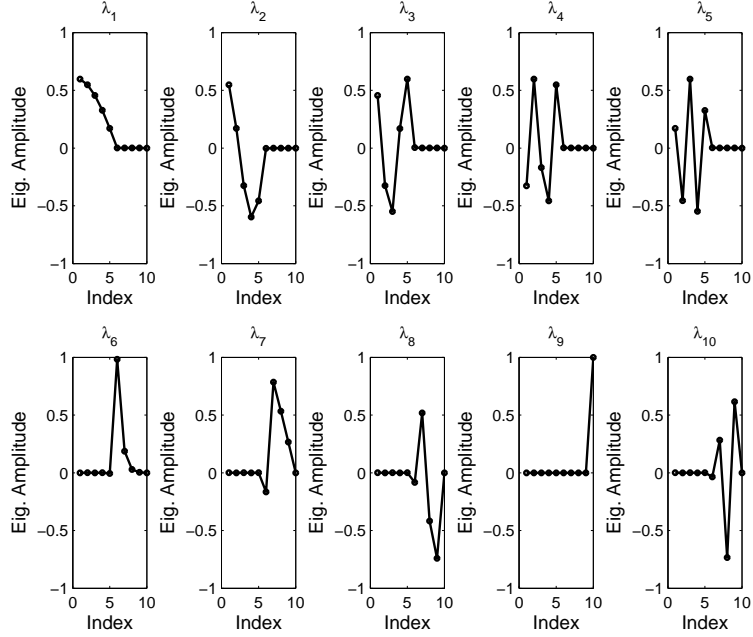}
\caption{Eigenvectors corresponding to stable solution at $F=2$.}
\label{fig:Auvectors}
\end{figure}

\begin{figure} [htbp]
\includegraphics[width=10truecm]{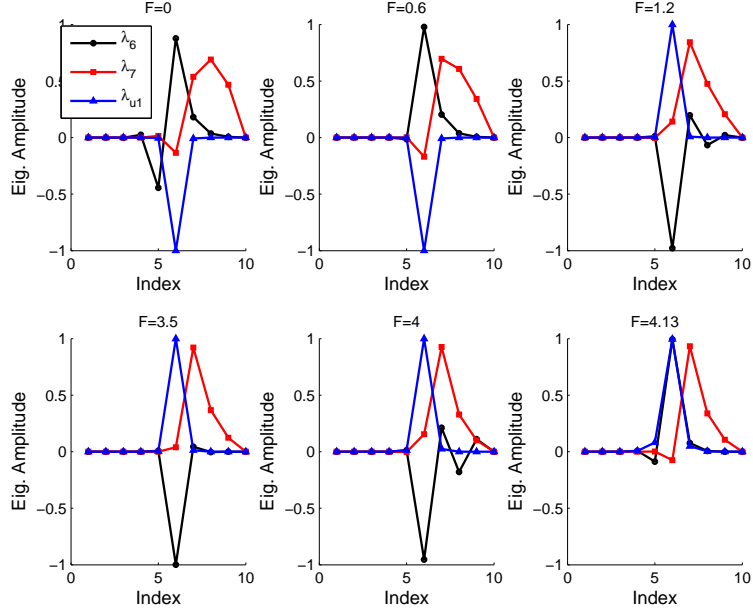}
\caption{(Color online) Eigenvectors corresponding to $\lambda_6$ and $\lambda_7$ of the stable solution and $\lambda_{u1}$ of the unstable solution at different forces.}
\label{fig:Auvectors_F}
\end{figure}

The eigenvectors corresponding to the stable domain wall solution at $F = 2$ are shown in figure \ref{fig:Auvectors} . A localized eigenvector at the $6^{th}$ bp appears while the others are spread over the remaining bps. The eigenvectors corresponding to eigenvalues from $\lambda_1$ to $\lambda_5$ are spread over the five sites located in the open side of the chain and have zero amplitude over the remaining sites. They describe the modes where the open bps oscillate. In a similar way, eigenvectors associated to $\lambda_7$ to $\lambda_{10}$ describe the oscillations of the closed side of the chain. The behavior of the eigenvectors corresponding to $\lambda_6$ and $\lambda_7$ (of the stable solution) and $\lambda_{u1}$ (of the unstable solution) at different forces are shown in figure \ref{fig:Auvectors_F}. The unstable solution eigenvector remains localized at $6^{th}$ bp for all values of $F$. The eigenvector corresponding to $\lambda_6$ becomes localized after $F>1.1$. Near the depinning force both eigenvectors coincide.

As the domain wall is very sharp, the positions of the frequency bands as well as the $F_d$ are almost the same if other sequence lengths and opening positions are used.

This method can be extended to a homogeneous CG chain. The results are similar but the oscillation modes are obtained at different frequency values and the depinning force is larger. 


\subsection{Estimation of the escape rates}
We can use these previous results to make some predictions on the thermal behavior of the unzipping process. One should be aware that the behavior of the domain wall solution can be approached by that of a single particle. This is because of the strong localization of such solution around the fork site \cite{Martinez1997, JJMazo2008}. Indeed, in the presence of thermal noise, the unzipping process can be regarded as the problem of the escape of a particle from a well \cite{Peyrad_2004}. This problem has been widely studied and some analytical expressions for the escape rate $k_e$ have been proposed at different damping regimes  \cite{Kramers,Haggin,mazo1013}. In the moderate to high damping regime the following equation can be used:

\begin{equation}
k_{e}=\{[1+(\frac{\gamma}{2\omega_b})^2]^{1/2}-\frac{\gamma}{2\omega_b}\}\frac{\omega_a}{2\pi}e^{-E/k_BT}
\label{eq:moderate}
\end{equation}
where $\omega_a$ and $\omega_b$ are related to the curvature at the well and the barrier respectively and $E$ is the energy of the barrier. From this formula $k_e$ is given in units of frequency (0.82 THz).

\begin{figure} [ht]
\includegraphics[width=10truecm]{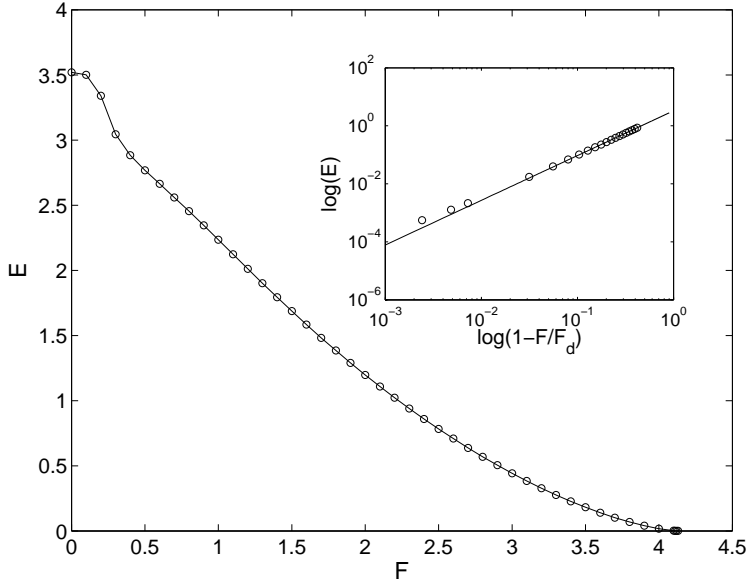}
\caption{Peierls-Nabarro energy at different forces. Inset: Fit of the $E$ vs $1-F/F_d$ near $F_d$.}
\label{fig:energy}
\end{figure}

In our case, the displacement of the domain wall corresponds to the escape of the particle from the well. $\omega_a$, $\omega_b$ and $E$ are the effective parameters of the well and the barrier that take into account the action of the remaining bps and the applied force. We take for $w_a$ the eigenvalue of the stable solution corresponding to the $6^{th}$ bp; for $w_b$ the eigenvalue of the unstable solution and for $E$ the height of the Peierls-Nabarro barrier which is the energy that the domain wall has to overcome to displace to the next bp site (which corresponds to the motion of $6^{th}$ bp). This term is given by $E=E_{unstable}-E_{stable}$, where $E_{unstable}$ is the energy of the unstable solution and $E_{stable}$ is that of the stable one for a given value of $F$, figure \ref{fig:energy}. As expected, the value of $E$ falls to zero at $F=F_d$. Near $F_d$, the dependence of $E$ with $F$ follows the relation $E\propto(1-F/F_d)^{1.5}$ as shown in the inset of \ref{fig:energy}. This is the expected behavior for a single particle inside a well \cite{Fulton1974}. That shows that the approximation of replacing the domain wall by a single particle should be good enough. The behavior of $\omega_a$ and $\omega_b$ as function of $F$ have been shown in figure \ref{fig:eigenvalues}.

\begin{figure}[htbp]
\includegraphics[angle=90,width=7truecm,angle=-90]{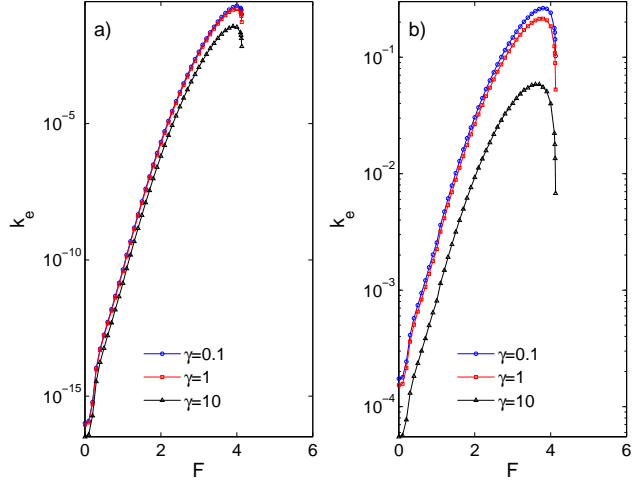}
\caption{(Color online) Escape rates at different damping and temperature values. Units of $k_e$ are those of the frequency (0.82 THz). a) $T=0.1$ b) $T=0.5$ .}
\label{fig:kramers}
\end{figure}

We estimate $k_e$ at different values of the applied force, damping and temperature (see figure \ref{fig:kramers}). At low temperature, $T=0.1$, $k_e$ is almost zero for forces below $F=1$. Beyond this force value, there is an increase on $k_e$ because of the decrease on the barrier energy. Close to $F_d$, the escape rate falls abruptly although energy barrier tends to zero. This is due to the abrupt decreasing of $\omega_a$ and $\omega_b$ at this point. However, in these ranges of the force, expression \ref{eq:moderate} is unphysical. One expects to have reliable results for barriers $E/k_BT>3$ \cite{JJMazo2010}. When temperature is increased, $T=0.5$ (room temperature), the behavior of $k_e$ with the force is similar. As expected, the values are larger than those for $T=0.1$. We look at the values of $k_e$ for $F=0.2$ (which is of the order of experimental forces) for both temperatures and $\gamma=0.1$. The obtained values are $k_e\sim10^{-16}$ and $k_e\sim10^{-4}$ for $T=0.1$ and $T=0.5$, respectively. This is the reason for which the value of $F_d$, defined at zero temperature, is high when compared with experiments. Indeed, DNA mechanical unzipping is an assisted thermal process dependent on the applied force. The values obtained for $k_e$ also suggest that other frequency dependent phenomena as the stochastic resonance or resonant activation would take place at frequencies lower than those obtained from the linear stability analysis.

\subsection{Influence of the ac field and temperature in the unzipping}

Finally, we study the influence of an ac external field and thermal noise on the unzipping process. The frequencies values of the field used here are in the interval of the modes obtained in the previous section. These values are in the order of $1 THz$. The ac field is applied over all bps of the chain. Two damping $\gamma=0.1$ and $\gamma=1$, and four temperature values are used. The field amplitude is $A=0.4$. This value is small when compared with the depinning force. The parameter values are in the same order of those used in \cite{Ana} when studying the influence of an ac field in the melting transition and bubble formation on DNA.

\begin{figure}[htbp]
\includegraphics[width=6.5truecm,angle=-90]{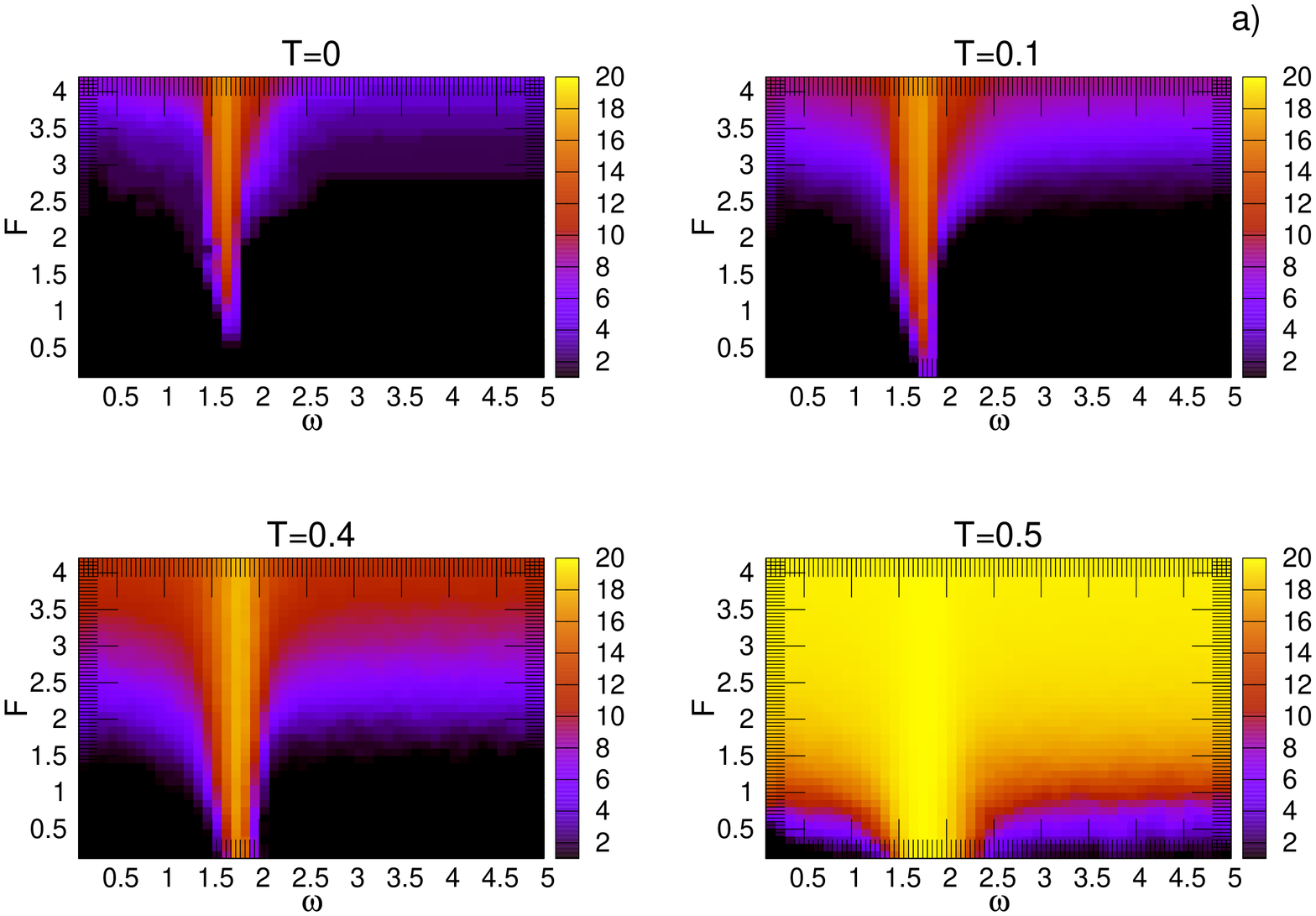}
\includegraphics[width=6.5truecm,angle=-90]{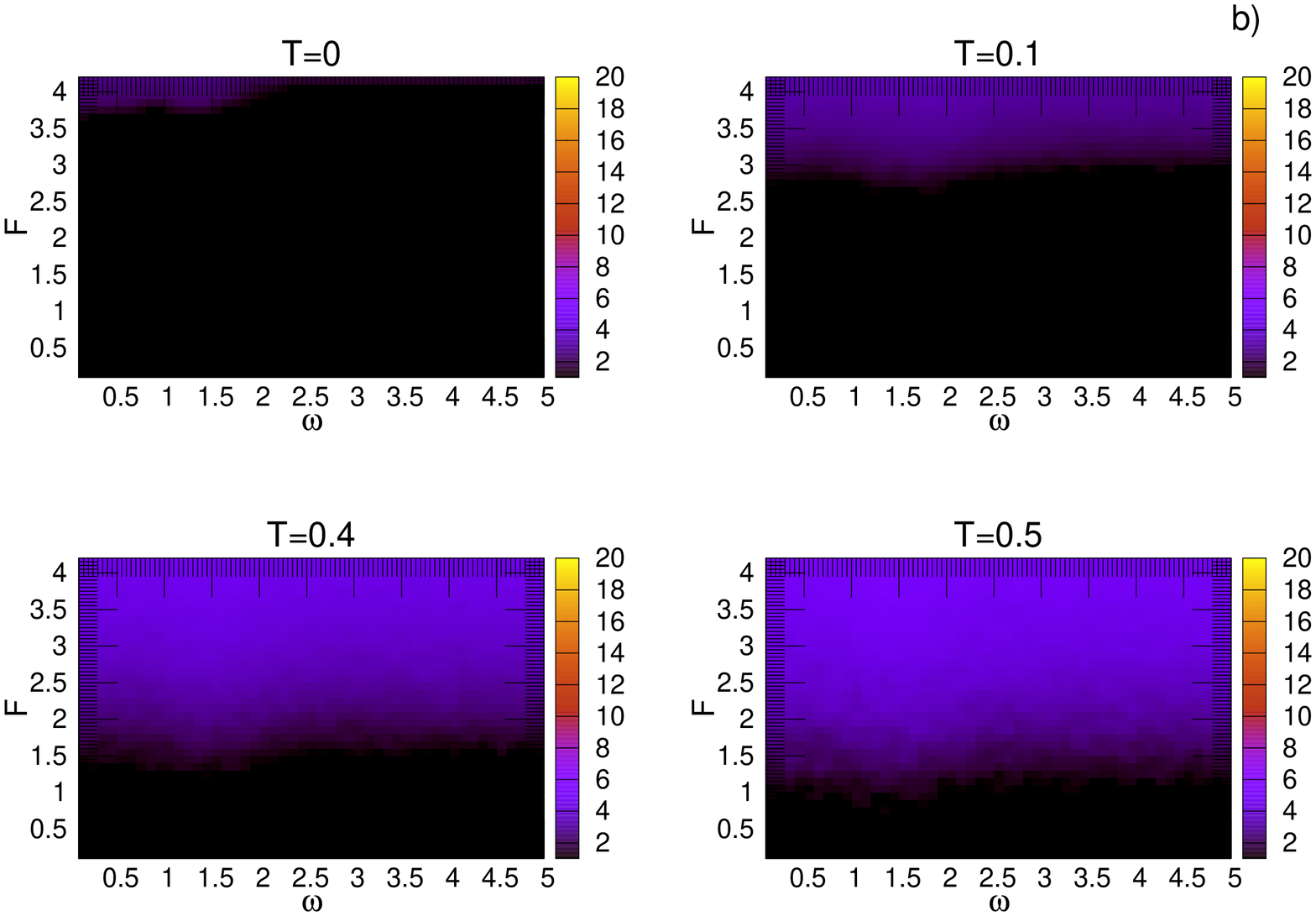}
\caption{(Color online) Influence of ac field on the domain wall position at different temperatures (see text). a) $\gamma=0.1$. b) $\gamma=1$. Note that now the simulated chain consists of 20 sites.}.
\label{fig:barr_frec}
\end{figure}

In these simulations we use a homogeneous AT chain of 20 bps and start from a closed state of the chain. The equations of motion are thermalized during a time $t_s=1000$ without force. After this time, both the constant force $F$ and the ac field $A$ are included in the motion equations and the system is integrated during $t_s=600$ (which corresponds approximately to 100 oscillation cycles of the external field). Then, we compute the temporal average of the domain wall position at different values of the frequency and the constant force $F$. $F$ is varied from $F=0.1$ until $F=F_d$. The results are averaged over 10 realizations. Figure \ref{fig:barr_frec} shows the mean domain wall position (on color scale) at different temperatures: $T=0, T=0.1 (60 K), T = 0.4 (240 K), T= 0.5 (300 K)$.

For $\gamma=0.1$ and $T=0$, an increase on domain wall position is obtained for frequency values in the range $1.4<\omega<1.8$ (a maximum of the mean domain wall position is observed around these frequency values). These values correspond to those of the upper frequency band obtained previously, i.e., to the oscillation modes of bps inside the well of the Morse potential. The peak is slightly asymmetric which may be explained for the presence of the oscillation mode of the $6^{th}$ bp. Thus, the mechanism of opening is resonant at these frequencies. As temperature increases the band spreads and the force for opening a given number of bp decrease because thermal noise assists opening events. This has been interpreted as a reduction of the critical force for unzipping due to resonance and temperature effects. For a larger damping $\gamma=1$, there is little influence of the frequency on the forces to unzipping the chain. This is the behavior expected for a resonant mechanism. These results are in agreement with \cite{Ana}, where maximum displacements were found around $\omega=9 rad/ps$ for a homogeneous AT chain and decreased when a higher damping was used. The oscillations modes corresponding to the open part of the chain do not influence the unzipping process. Simulations were also carried out at different field amplitudes (results not shown). In agreement with \cite{Ana}, the resonant influence of the field is observed for field amplitudes larger than a given threshold.

\begin{figure}[htbp]
\includegraphics[width=6.5truecm,angle=-90]{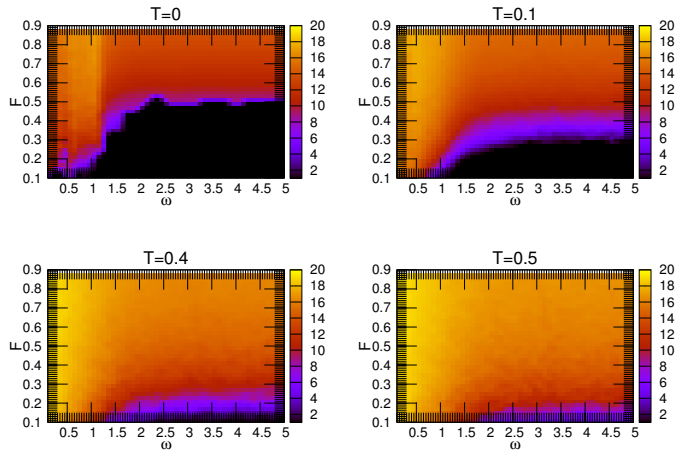}
\caption{(Color online) Influence of ac field on the domain wall position at different temperatures in the non barrier case and $\gamma=0.1$. Note that now the simulated chain consists of 20 sites. Note also that force scale is different to than that used in \ref{fig:barr_frec}.}
\label{fig:nobarr_frec}
\end{figure}

It is important to point out that the resonant influence of the ac field in the unzipping process, as well as in the melting and the bubble formation in the framework of the PBD model, is only possible with the inclusion of the barrier term in the standard Morse potential.  When this term is absent, there is no resonant peak in the dependence on the frequency of the external ac field, at least in the interval we have studied. This is mainly due to the stability that the barrier confer to the domain wall solution. As we have seen, a frequency can be associated to the domain, that one corresponding to the localized eigenvector at the fork position. This frequency is in the range of the upper band of the linear stability spectrum. So, the domain wall frequency has no sense without the barrier and we do not expect to have resonant behavior. In order to check this prediction, we have performed a numerical simulation of the model without the barrier in the same range of parameters. Figure \ref{fig:nobarr_frec} shows the result for the same parameter values as in figure \ref{fig:barr_frec}. Also, due to the lack of a barrier, the depinning force is reduced. Anyway, one can observe a drastic change in the behavior respect to the frequency. No resonant reduction of the unzipping force is revealed.

Thus, the inclusion of this term not only produces some results closer to experimental ones (sharper melting transition, wider and with larger lifetime bubbles) \cite{Weber2006,Peyrard2009,Tapia} but also leads to new features that are absent in the model without the inclusion of the barrier.

\section{CONCLUDING REMARKS}
In summary, we have studied the features of the unzipping process when a Gaussian barrier is added to the standard Morse potential and the influence of an external alternating field on the dynamics of this process. We have obtained the linear frequency modes corresponding to different opening states of the chain. The frequencies that influence resonantly in the unzipping process are those of the close part of the chain. We were also able to estimate the escape rates for the unzipping process under different parameters (temperature and damping). The knowledge of this parameter is important to understand force spectroscopy experiments and other frequency dependent mechanisms as stochastic resonance and resonant activation. The analysis we have presented here may be extended to real chains if  sequence dependent parameters are included. The inclusion of the barrier in the Morse potential leads to new features that are absent in the original PBD model with regard to the response to an ac external field. It should be stressed that also the inclusion of external fields provokes a dramatic reduction of the critical unzipping force at resonant frequencies. This could give a way for testing the validity of the inclusion of this term in a real experiment where this mechanism can be checked.

\begin{acknowledgments}
We thank J.J. Mazo for discussion and critical reading of the manuscript. This work is supported by the Spanish DGICYT Projects No. FIS2011-25167, co-financed by FEDER funds, and by the Comunidad de Arag\'on through a grant to the FENOL group. AEBP acknowledges the financial support of University of Zaragoza and Banco Santander.
\end{acknowledgments}


\begin{thebibliography}{99}

\bibitem{Bockelmann1997} B. Essevaz-Roulet, U. Bockelmann, and F. Heslot, Proc. Natl. Acad. Sci. USA 94 (1997) 11935.

\bibitem{Bockelmann2002} U. Bockelmann, Ph. Thomen, B. Essevaz-Roulet, V. Viasnoff, and F. Heslot, Biophysical Journal 82 (2002) 1537.

\bibitem{Danilowicz2003} C. Danilowicz, V. W. Coljee, C. Bouzigues, D. K. Lubensky, D. R. Nelson, and M. Prentiss, Proc. Natl. Acad. Sci. USA 100 (2003) 1694.

\bibitem{Ritort2010} J. M. Huguet, C. V. Bizarro, N. Forns., S. B. Smith, C. Bustamante., and F. Ritort., Proc. Natl. Acad. Sci. USA 107 (2010) 15431.

\bibitem{Alex_PRA} B. S. Alexandrov, V. Gelev, A. R. Bishop, A. Usheva, and K. O. Rasmussen, Phys. Lett. A 374 (2010) 1214.

\bibitem{Alex_PRE} P. Maniadis, B. S. Alexandrov, A. R. Bishop, and K. O. Rasmussen, Phys. Rev. E 83 (2011) 011904.

\bibitem{Ana} A. E. Bergues-Pupo, J. M. Bergues, and F. Falo, Phys. Rev. E 87 (2013) 022703.

\bibitem{Alex_PLoSCB} J. Bock, Y. Fukuyo, S. Kang, M. L. Phipps, L. B. Alexandrov, K. O. Rasmussen, A. R. Bishop, E. D. Rosen, J. S. Martinez, H. T. Chen, G. Rodriguez, B. S. Alexandrov, and A. Usheva, PLoS ONE  5 (2010) 15806.

\bibitem{Bruan_2004} O. Braun, A. Hanke, and U. Seifert, Phys.Rev.Lett. 93 (2004) 158105-1.

\bibitem{Ritort2012} K. Hayashi, S. de Lorenzo, M. Manosas, J. M. Huguet and F. Ritort, Phys. Rev. X 2 (2012) 031012.


\bibitem{Kumar2013} G. Mishra, P. Sadhukhan, S. M. Bhattacharjee, and S. Kumar, Phys. Rev. E 87, (2013) 022718.

\bibitem{Peyrad1993} T. Dauxois, M. Peyrard, and A. R. Bishop, Phys. Rev. E  47 (1993) 684.

\bibitem{Santiago2005} S. Cuesta-Lopez, J. Errami, F. Falo, and M. Peyrard, J. of Biol. Phys. 31 (2005) 273.

\bibitem{Voulgarakis2006} N. K. Voulgarakis, A. Redondo, A. R. Bishop, and K. O. Rasmussen, Nano Letters 6 (2006) 1483.

\bibitem{Voulgarakis2005} N. K. Voulgarakis, A. Redondo, A. R. Bishop, and K. O. Rasmussen, Phys. Rev. Lett. 96 (2006) 248101.

\bibitem{Singh2013} A. Singh, B. Mittal, and N. Singh, Physics Express, 3 (2013) 18.

\bibitem{Webber2009} G. Weber, J. W. Essex, and C. Neylon, Nat. Phys. 5 (2009) 769.

\bibitem{Weber2006} G. Weber, Europhys. Lett. 73 (2006) 806.

\bibitem{Peyrard2009} M. Peyrard, S. Cuesta-Lopez,  G. James, J. Biol. Phys. 35 (2009) 73.

\bibitem{Tapia} R. Tapia-Rojo, J. J. Mazo, and F. Falo, Phys. Rev. E 82 (2010) 031916.

\bibitem{Peyrad_2004} M. Peyrard, Nonlinearity 17 (2004) R1.

\bibitem{sde2} H. S. Greenside and E. Helfand, The Bell System Technical Journal 60 (1981) 1927.

\bibitem{sde1} E. Helfand, Bell System Technical Journal 58 (1979) 2289.

\bibitem{Strogatz} S. H. Strogatz, {\emph Non Linear Dynamics an Chaos}, Perseus Books (1994) Reading, Massachusetts.

\bibitem{Kramers} H. Kramers, Physica 7 (1940) 284.

\bibitem{Haggin} P. Hanggi, P. Talkner, and M. Borkovec, Rev. Mod. Phys. 62 (1990) 251.

\bibitem{Martinez1997} P.J. Martinez, F. Falo, J.J. Mazo, L.M. Floria and A. Sanchez
Phys. Rev. B 56  (1997) 87.

\bibitem{JJMazo2008} J. J. Mazo, F. Naranjo, and K. Segall, Phys. Rev. B, 78 (2008) 174510.

\bibitem{mazo1013} J.J. Mazo, O.Y. Fajardo, and D. Zueco, J. Chem. Phys. 138 (2013) 104105.

\bibitem{Fulton1974} T. A. Fulton, and L. N. Dunkleberger, Phys. Rev. B 9 (1974) 4760.

\bibitem{JJMazo2010} J.J. Mazo, F. Naranjo, and D. Zueco, Phys. Rev. B 82 (2010) 094505.



\end{thebibliography}
\end{document}